\documentclass[12pt]{iopart}
\usepackage{graphicx} 
\usepackage[numbers]{natbib}
\usepackage{amsmath}

\usepackage{multirow}
\usepackage{subcaption}
\usepackage{hyperref}

\usepackage{ulem}
\usepackage{xcolor}
\usepackage{csquotes}

\begin{document}

\title[]{Analysis of flows in social media uncovers a new multi-step model of information spread}

\author{Matteo Serafino $^1$, G. Virginio Clemente$^{1,2}$, James Flamino$^3$, Boleslaw K. Szymanski$^{3,4}$, Omar Lizardo $^5$, Hern\'an A. Makse$^1$}

\address{$^1$ Levich Institute and Physics Department, City College of New York, New York, NY  10031, USA}
\address{$^2$IMT School for Advanced Studies, 55100 Lucca, Italy}
\address{$^3$Department of Computer Science and NEST Center, RPI, Rensselaer Polytechnic Institute, Troy, NY, USA}
\address{$^4$ Spo\l{}eczna Akademia Nauk, \L{}\'{o}d\'{z}, Poland}
\address{$^5$ University of California, Los Angeles, CA, USA}

\ead{matteo.serafino1991@gmail.com}

\begin{abstract}
Since the advent of the internet, communication paradigms have continuously evolved,
resulting in a present-day landscape where the dynamics of information dissemination
have undergone a complete transformation compared to the past. In this study, we challenge
the conventional two-step flow model of communication, a long-standing paradigm in the field.
Our approach introduces a more intricate multi-step and multi-actor model that effectively
captures the complexities of modern information spread. We test our hypothesis
by examining the spread of information on the Twitter platform. Our findings support
the multi-step and multi-actor model hypothesis. In this framework,
influencers (individuals with a significant presence in social media)
emerges as new central figures and partially take on
the role previously attributed to opinion leaders. However, this does not
apply to opinion leaders who adapt and reaffirm their influential
position on social media, here defined as opinion-leading influencers.
Additionally, we  note a substantial number of adopters directly accessing
information sources, suggesting a potentialdecline if influence
in both opinion leaders and influencers. Finally, we found distinctions
in the diffusion patterns of left- and right-leaning groups,
indicating variations in the underlying structure of information
dissemination across different ideologies.

\end{abstract}
\noindent{\it Keywords}: Information diffusion, Influencers, Opinion leaders

\submitto{Journal of Statistical Mechanics: Theory and Experiment}
\maketitle

\section{Introduction}

\begin{figure*}
\centering
  \begin{subfigure}[b]{0.35\textwidth}
    \centering
    \includegraphics[height=2in]{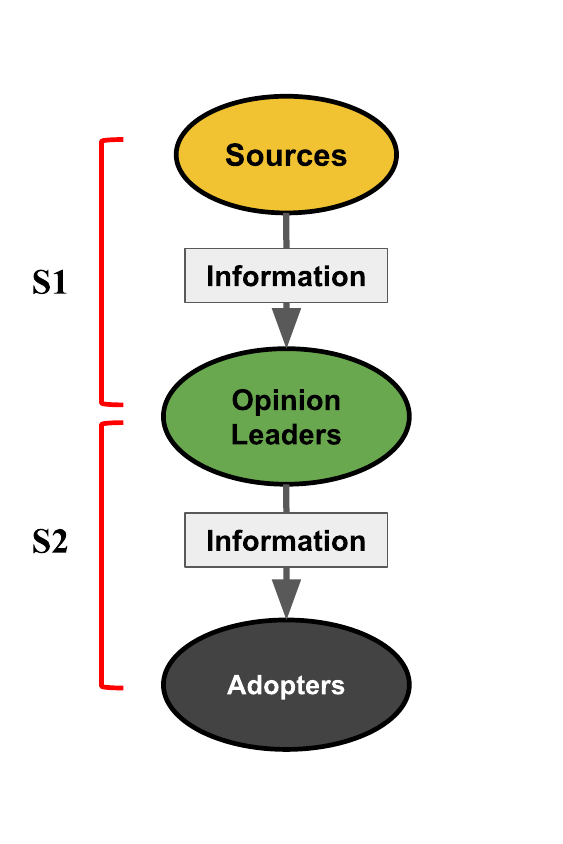}
    \caption{}
  \end{subfigure}
  \hfill
  \begin{subfigure}[b]{0.60\textwidth}
    \centering
    \includegraphics[height=2in]{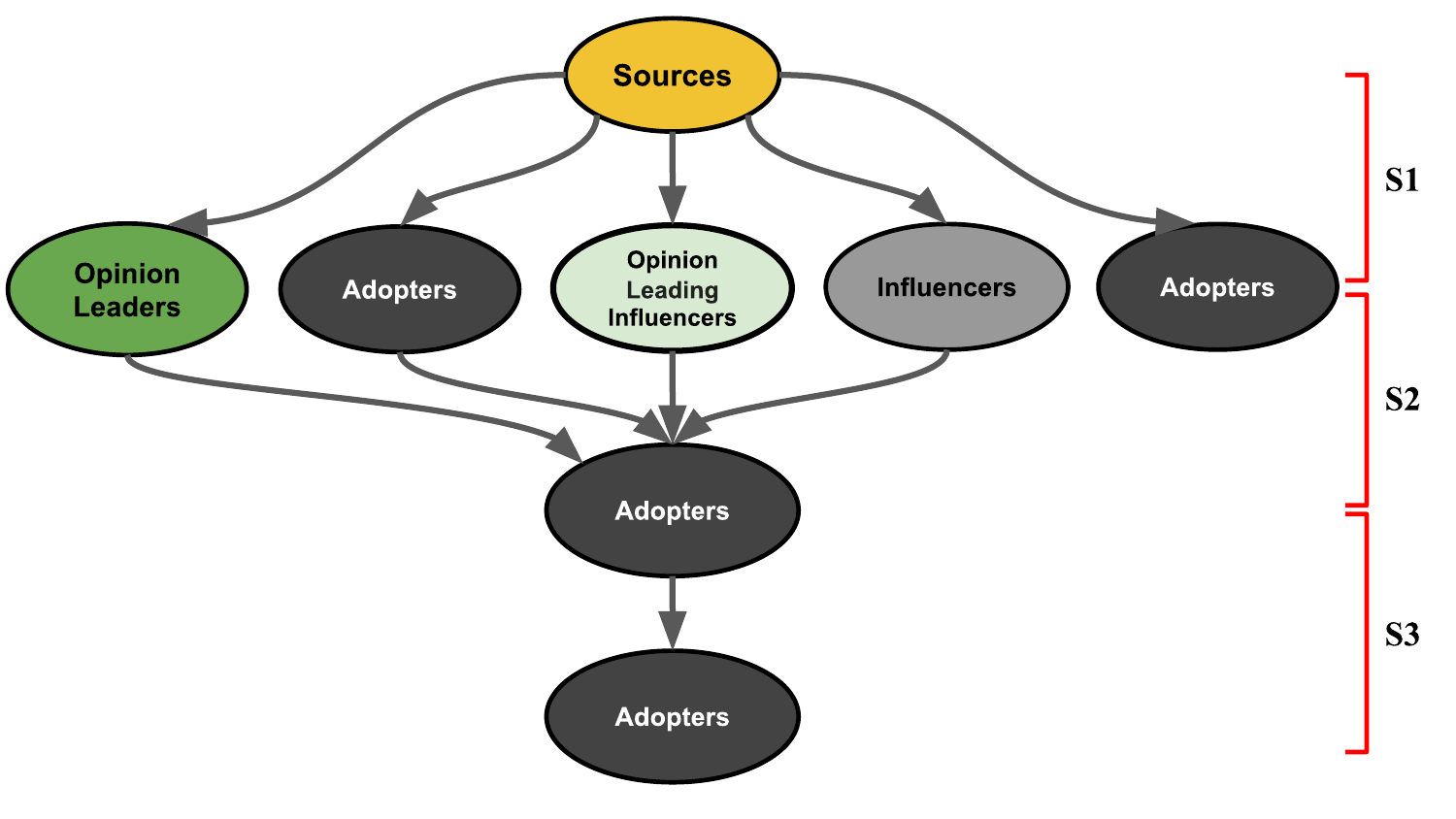}
    \caption{}
  \end{subfigure}
\caption{\textbf{Two Information Diffusion Models.} (a) The traditional
two-step model of information diffusion involves the flow of information
to adopters through the mediation of opinion leaders. (b) The multistep
model for information diffusion. Adopters can directly access the original
information or obtain it through the more traditional opinion leaders
or influencers. Note the possibility of \enquote{horizontal information flow},
where adopters receive information from other adopters.}
\label{fig:figure1}
\end{figure*}

The rise of social media has led to a fundamental transformation
in how information, opinions, and beliefs propagate in contemporary society.
Traditional models of information diffusion developed in Sociology and
Communication in the mid-twentieth century  \cite{katz1955, katz1957,
lazarsfeld1944, berelson1954, weimann1982} presupposed a linear \enquote{two-step}
flow of information from sources to the mass public mediated by \enquote{opinion leaders}
\cite{lazarsfeld1944, burt1999, watts2007, rogers1962}, defined as recognized experts or respected public figures with acknowledged credibility
in specific fields (see Fig. \ref{fig:figure1}a). For example, within
the two step model (Fig. \ref{fig:figure1}a), The New York Times (the source)
releases a news article on the evolution of $R_{0}$, a key epidemiological
metric for measuring infectious agent transmissibility.  Dr. Anthony Fauci,
acting as an opinion leader, reads and simplifies the news (S1: first step) 
before disseminating it to the broader population in a more accessible manner. These individuals, termed adopters, engage with or adopt the idea at various stages (S2: second step).

In the current scene, by way of contrast, information diffuses via
more complex \textit{multi-step} flows, including one-step flows with
adopters accessing information directly from the sources \cite{bennett2006},
without intermediaries, traditionally mediated two-step, and more complex,
longer-path dynamics featuring a heterogeneous set of agents. For instance,
adopters may obtain information from other adopters,  who, in turn, receive
the information from an opinion leader or other mediators (see Fig. \ref{fig:figure1}b).
Notably, a new figure has emerged within the intricate structure of contemporary
digitally-mediated information diffusion: the \textit{influencer}. Unlike
traditional opinion leaders, influencers often build their authority through
a combination of relatability, engaging content, and a substantial online
presence \cite{casalo2020influencers}. 

The rise of influencers has added a new layer to the landscape of
information diffusion, introducing a dynamic where individuals
with significant followings can swiftly impact trends and opinions.
In the context of COVID-19, Elon Musk can be considered a notable
influencer who wields considerable social influence that can shape
public opinions. Musk's impact is not necessarily rooted in expertise
in infectious diseases or pandemics but rather in his extensive reach
across online social platforms. Of course, opinion leaders and influencers
need not be mutually exclusive groups. Instances exist where traditional
opinion leaders have effectively established themselves as influencers.
We refer to these individuals as opinion-leading influencers.
A notable example is Helen Branswell, a Canadian infectious diseases and
global health reporter at Stat News. With a fifteen-year tenure as a medical
reporter at The Canadian Press,  she spearheaded Ebola, Zika, SARS, and swine
flu pandemics coverage. Beyond her field \textcolor{blue}{of} expertise, Branswell maintains a
robust online presence, qualifying her as an opinion-leading influencer. 

While most observers agree that the traditional opinion-leader-mediated two-step dynamic
has certainly been disrupted \cite{bennett2006,karlsen2015, wu2011,
weeks2017, winter2021, choi2015, dubois2014, turcotte2015}, we know little about the new
pathways and actors through which information diffuses
in online social media, as well as the extent to which opinion leaders, influencers, 
or the new hybrid figure of the opinion-leading influencer still serve as key mediators.
Or whether individuals have become direct consumers of information from sources,
or the extent to which horizontal transmission among adopters,
defined here as person-to-person transmission, in contrast
to the top-down (vertical) transmission from opinion leaders to individuals,
accounts for the bulk of information flow. Despite considerable speculation about
how social media has transformed information diffusion, there is still a need for
quantitative studies that allow us to clarify from different perspectives the
extent to which information flow on digital platforms is mediated through
multiple steps. This paper sets out to
reconstruct the pathways through which information flows
in the era of social media, to characterize how information
diffuses through different groups of actors and to ascertain whether
the decline in the influence of opinion leaders
\cite{bennett2006,bennett2008new} has been
greatly exaggerated or not.

For this task, we turn to Twitter content, using a dataset centered
around the US 2020 presidential elections \cite{flamino2023political}.
Digital platforms like Twitter provide researchers with the tools to
track specific Uniform Resource Locators (URLs) released by the sources.
This tracking unveils insights into users who share the same URL,
offering a detailed account of user interactions over time. This method
enables us to directly assess a proxy for the channels of information flow among
various actors, encompassing the source, opinion leaders, influencers, 
opinion-leading influencers, and ultimately, adopters. 
We leverage this distinctive capability to disclose the structural
characteristics and dynamics initiated by various actors in the digital space. 

\section{Results}
Our analysis proceeds as follows (also refer to  Appendix, Fig. A1).
Initially, from the raw data, we construct the retweet network by considering
tweets that include a URL linking to one of the news outlets listed in Appendix,
Table A1. Considering only tweets with URLs allows us to trace
back the original source of the information, a key aspect in
understanding the diffusion information process explained in the introduction. 
In the retweet network, a connection exists from
user $i$ to user $j$ if $j$ has retweeted a tweet from $i$. 
The connection is weighted by the number of times $j$ has
retweeted a tweet from $i$. Additionally, we assign an average
retweet time to each link, calculated as the average retweet time among 
the retweets of $i$ by $j$. However, this process results in the loss
of the temporal dimension of the interactions. We then validate these links against an
entropy-based null model. This validation process is designed to
preserve only significant connections, thereby reducing potential
biases in subsequent stages of our methodology. This step is essential
due to the original structure of the data, which does not allow
for direct measurement of the patterns of interest (Section \ref{MaterialandMethods}).
Next, using the Collective Influence (CI) algorithm (see Appendix, Section B), we identify
the top 1000 influencers, representing the 0.1\% of the users in the network, 
and accounting for more than 65\% of the total connections. We classify them
into one of the following categories: opinion leaders, influencers,
opinion-leading influencers, adopters, and sources (Refer
to Section \ref{Influencers and Opinion Leaders Identification} and
Table \ref{tab:description_actors} for more information on the
classification process). It is important to note that, while the choice of considering
the top 1000 influencers according to CI is arbitrary, increasing
this number does not significantly alter the final results of our analyses.
This is because the remaining nodes (which account for more than 99\% of
the total nodes) contribute to less than 35\% of the remaining links.
Finally, we apply the Breadth-First Search (BFS) algorithm on the
validated network to uncover the underlying structure most likely
to facilitate information propagation. In this context, 'most likely'
refers to the frequency of retweets between groups of nodes, without
accounting for the temporal dimension, which is not preserved when
constructing the retweet network. We refer to this extracted structure
as the 'backbone.' It represents the skeleton of information diffusion,
meaning that if a news is shared on Twitter, it would most likely
spread through the identified skeleton (or one of its subgraphs). 
As detailed in Appendix A.1, the results remain robust when filtering
the links based on their average retweet time. Further analysis, 
provided in Appendix C, contributes to validating the results.
Finally, we leverage the tendency of each news outlet to exhibit a bias
toward either a left or right ideology. This enables a more in-depth analysis
of the diffusion structure and highlights potential differences between
diverse ideological perspectives.

\subsection{Modeling Information Flow}
Here, our emphasis is on reconstructing the flow of information 
diffusion originating from the sources documented in the
Appendix, Table A1, regardless of the news outlet bias under
consideration. The resulting retweet network consists of 2 963 210
nodes and  27 608 480 unique connections. The link validation
procedure validates 51\% of the total links, with the validated
network consisting of 1 775 194 nodes and 14 258 411 connections.
The decrease in the number of nodes is because some nodes
are isolated after validation and, therefore, discarded.
In this graph, we identify 1 173 opinion leaders (politicians
or users affiliated with the journal under consideration, see
Appendix, Table A1), 241 sources, 520 opinion-leading influencers, 
399 influencers, and 1 772 869 adopters. Refer to
Section 4.4 of Material and Methods for a better understating on
the classification, and to the section Data, Materials,
and Software Availability for a comprehensive list of classified
sources and opinion leaders.
We consider influencers the users within the top 1000 users
by CI who are not opinion leaders. 
The latter indeed are the opinion-leading influencers.
Although, in theory, the sum of these two categories should total 1000, 
practical deviations occur due to the inclusion of some sources,
which also fall within the top 1000 users by CI.
Moreover, notice here that despite the number of the initial
news outlets consists of 69 elements, here when
we refer to sources we consider all the accounts
associated with one of those news outlets. For example, CNN is
associated with \/@CNN but also with \/@CNNPolitics. 
Table A2 in the Appendix presents the top 10 influencers,
opinion-leading influencers, and opinion leaders.

\begin{figure*}[ht]
\centering\includegraphics[width=0.75\textwidth]{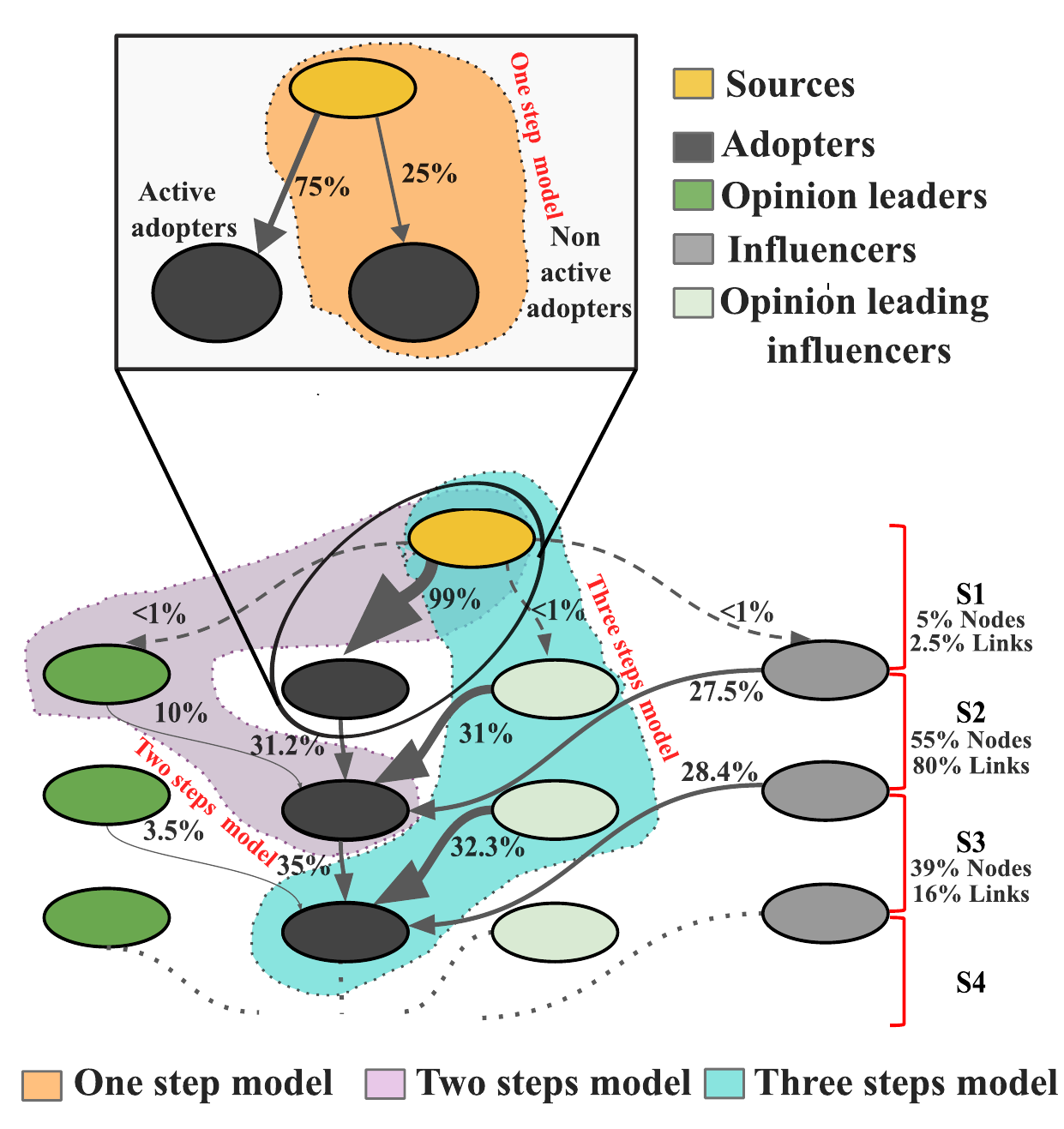}
    \caption{{\bf Information flow backbone}.Skeleton of the information flow. We define active adopters as individuals directly involved in subsequent steps of the flow, in contrast to Non active adopters.
    The steps of information flow are highlighted on the right, indicating steps (S1, S2, S3, S4) along with the percentage of corresponding node and connection counts. Overlapping groups are non-existent within the same step and between different steps. Step one (S1) primarily involves mediators (opinion leaders, influencers, opinion-leading influencers, and adopters) retweeting the sources. Step two (S2) consists of adopters retweeting mediators from S1. This pattern is repeated until S9. The final 6 steps make up the remaining 1\% of nodes and 1.5\% of the links in the skeleton. Not all steps are displayed for clarity, with an incomplete step S4 alluding to its continuation. Connection normalization per step ensures that the percentage on each layer adds up to 100. Throughout the paper, consistent color associations for the main actors are maintained.}
    \label{fig:multistep_full}
\end{figure*}

After identifying the main actors, we proceed to unveil
the skeleton of the information flow using the Breadth-First
Search (BFS) algorithm, (see Materials and Methods \ref{MaterialandMethods}). 
Apart from the final network structure, to support the implementation
of this procedure, we conducted a robustness check to examine the
connections directed towards the adopters (See Appendix C). These
connections can be classified in two ways in relation to the BFS-derived
structure. The first classification pertains to connections that contradict
the directions identified by the BFS, and the second relates to connections
that may link from step 1 to step 2 in the identified structure. In both cases,
however, we can assume that the impact of such connections is negligible,
as they account for much less than 1\% of the total connections.
The outcome is illustrated in Fig. \ref{fig:multistep_full}.
The normalization of connections
is computed per step, ensuring that the percentage
of connections in each step adds up to 100\%. The resulting
skeleton comprises 1 718 201 nodes and 5 306 961 links.

In the initial step (S1), a substantial group of adopters
directly access information from the sources without any mediator.
Less than 1\% of the connections in this step are directed to opinion
leaders, opinion-leading influencers, and influencers (dashed
arrows in the figure). Indeed adopters in this step (S1)
account for more than 99\% of the nodes accessing the
information directly from the sources. Overall this step 
accounts for 5\% of the total nodes and 2.5\% of the connections
in the skeleton found employing the BFS algorithm.
The significant difference between adopters and the other main
actors arises from the substantial size gap, with the adopter
group being tens of times larger than the other three groups,
each of which is of the same order of magnitude. Within the adopters 
identified in S1, 75\% are active adopters meaning they are directly involved in subsequent steps of the flow and have retweeted in S2, 
as indicated in the inset of Fig. \ref{fig:multistep_full}. 
Moreover, 25\%of the adopters (non active adopters) defined in S1
serve as information sinks, conforming to a one-step model
structure for information diffusion, as illustrated by
the orange cloud in the inset.

As the diffusion process progresses (S2 in Fig. \ref{fig:multistep_full}),
55\% of the nodes (80\% of the total links) in the skeleton access
the information through mediators. Traditional opinion leaders only account for
10\% of the connections, indicating a significant lower influence compared to other groups. 
opinion-leading influencers, on the other hand, due to their
adaptability in the online community, still wield strong influence,
accounting for 31\% of the connections. Similarly,
influencers position themselves with significant influence,
mediating 27.5\% of the connections. These results are
consistent with a two-step model structure
for information diffusion where mediators encompass
not only traditional opinion leaders (indicated by 
the magenta cloud in Fig. \ref{fig:multistep_full}) as
originally formulated, but a diverse set of actors, 
including influencers and opinion-leading influencers.
We observe three additional ways to construct a two-step model
by substituting opinion leaders with opinion-leading influencers,
influencers, or adopters. This forms the basis of a multi-actor model. 
Additionally, adopters also function as mediators, facilitating
the information transfer to other adopters (\enquote{horizontal information flow})
and accounting for 31.2\% of the connections in S2.

As depicted in Fig. \ref{fig:multistep_full},
the information diffusion extends beyond S2. The structure
presented in the third step (S3) is similar to the one in S2
and accounts only for 39\% of the nodes in the skeleton (16\% of the links).
The cyan cloud in Fig. \ref{fig:multistep_full} represents one of the four
potential three-step-like structures for information diffusion.
Importantly, from S2 onward, only links representing at least 1\%
of the connections in each step are displayed for visual clarity.
Therefore, connections between the opinion-leading influencers
between S2 and S3 in the cyan cloud exist, even if not explicitly shown.

This pattern is repeated until S9, with the remaining 6 steps
comprising the remaining 1\% of the nodes and 1.5\% of the connections
in the skeleton.  We do not display all the steps in
Fig. \ref{fig:multistep_full} for the sake of clarity.
However, we leave an incomplete step S4 to allude to its continuation.

\subsection{Left vs. Right}
In this section, we investigate whether the information
flow linked to various political media biases displays
distinct characteristics. Specifically, our
focus is on identifying potential differences between
content related to the left and right political
spectrum. We aim to uncover key disparities in both
the structure of information flow and the roles played
by primary actors in information dissemination.
Our analysis focuses exclusively on left-leaning sources (left
and left-leaning in Appendix, Table A1) for
studying the left, and right-leaning (right and right-leaning in
in Appendix, Table A1) news outlets for investigating the right.

By following the same steps as in the previous analysis
we obtain a skeleton comprising 475 636 nodes and 902 189 edges
for the right-leaning. This represents a reduction of 58\%
of the nodes and 87\% of the edges present in the original
retweet network. In the case of the left-leaning sources,
we end up with a skeleton consisting of 710 432 nodes and 
2 027 887 edges, indicating a reduction of 71\% of the nodes
and 90\% of the edges present in the original left retweet
network.

\begin{figure*}[ht!]
    \centering
    \begin{subfigure}{0.45\textwidth}
    \centering
        \includegraphics[width=\linewidth]{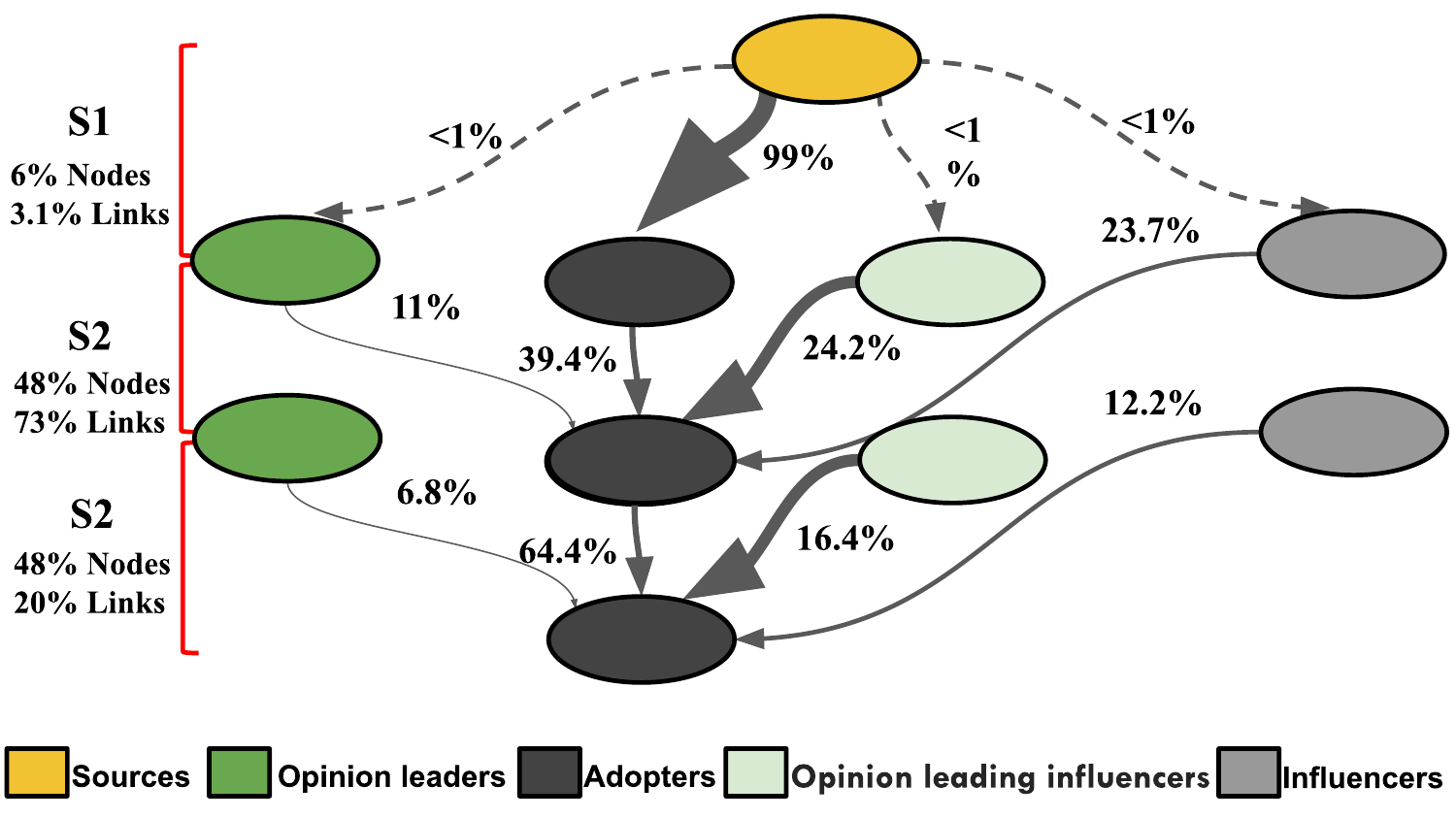}
        \caption{}
    \end{subfigure}
    \hfill  
    \begin{subfigure}{0.45\textwidth}
    \centering
         \includegraphics[width=\linewidth]{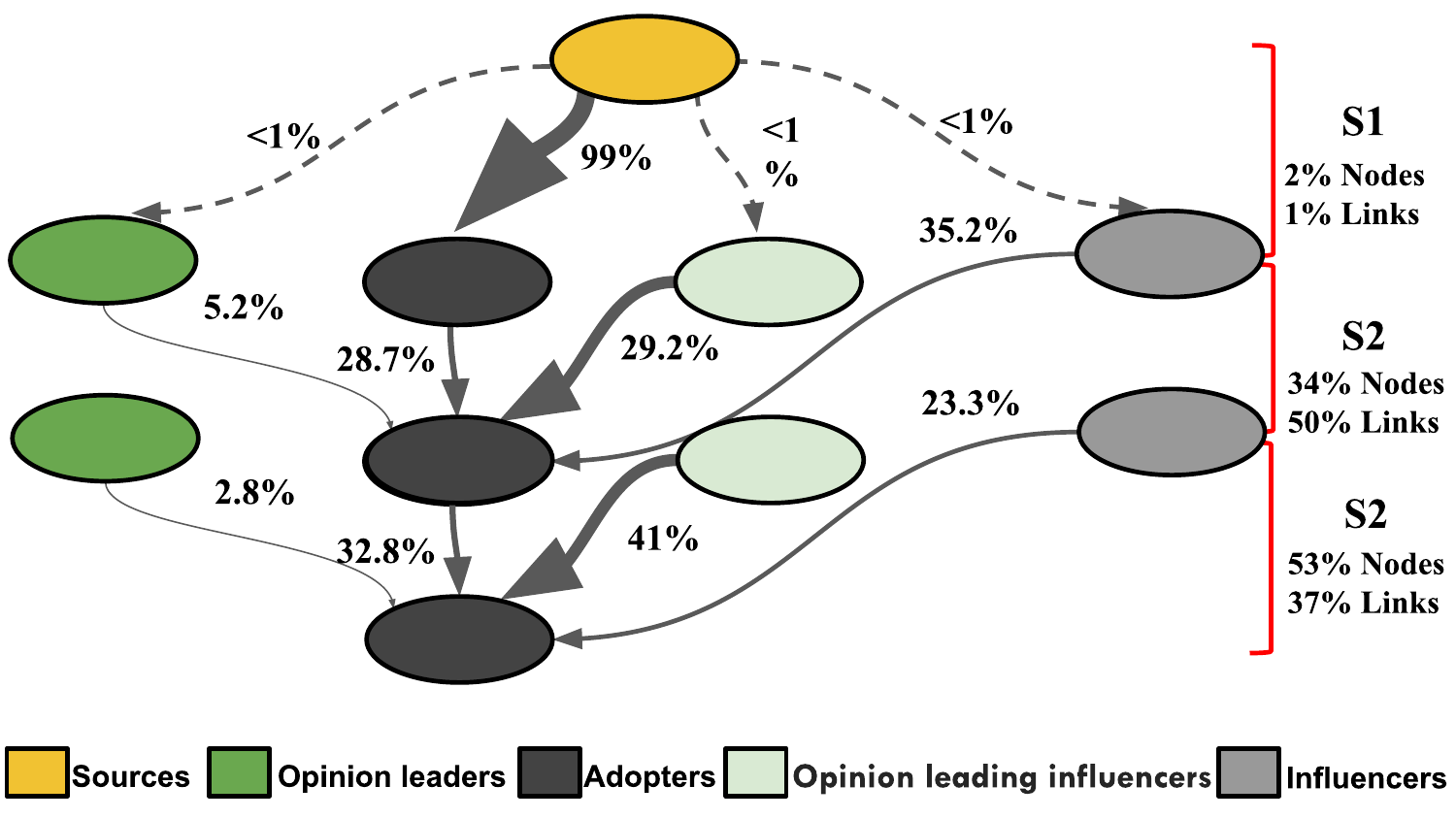}
        \caption{}
    \end{subfigure}
    \caption{ {\bf Backbone: Left vs. Right.} (a) Backbone obtained by following
    the steps depicted in Appendix, Fig. S2 but by only considering news coming
    from left and left-leaning news outlets. (b) Backbone obtained by following
    the steps depicted in Appendix, Fig. S2 but by only considering news coming
    from right and right-leaning news outlets. While stopping at the third step
    of the information flow, we confirm that the patterns we observed in the
    the general case still hold. Throughout the paper, consistent color associations
    for the main actors are maintained, as indicated at the bottom of the figure.}
    \label{fig:left_vs_right} 
\end{figure*}

Figure \ref{fig:left_vs_right} displays the backbones resulting by
considering as sources of information only the left-leaning news
outlets (on the left) and the right-leaning news outlet (on the right).
The figures stop at the third step of the information flow, 
which accounts for more than 96\% of the nodes and 95\% of the links
in the skeleton (in both cases). The structure of the consecutive layers
is similar to the one shown in Fig. \ref{fig:multistep_full}, in agreement 
with the hypothesis of a multi steps and multi actors model for the diffusion
of information, despite the political polarization of the information 
circulating. 

The initial distinction observed between left and right lies in
the depth of the information diffusion process. In the case of the left-leaning sources,
the first two steps of this process encompass nearly 55\% of the nodes and 76\%
of the connections. In contrast, these percentages decrease to 36\% and 51\%
for the right-leaning sources, as illustrated in Fig. \ref{fig:left_vs_right}.

Another intriguing difference highlighted in Fig. \ref{fig:left_vs_right} pertains
to the role of influencers in the second step (S2) of the information diffusion process. 
Specifically, our analysis reveals that, in the case of the right-leaning backbone, 35.2\% 
of the connections in S2 originate from adopters who retweeted influencers. 
In contrast, opinion leaders account for only 5.2\% of the connections, 
while opinion-leading influencers play a substantial role, being
retweeted 29.2\% of the time. Adopters rank third in terms of the
frequency with which they are retweeted, following opinion-leading influencers
and influencers.

This hierarchical pattern undergoes a shift in the left-leaning backbone.
Here, opinion-leading influencers are retweeted 24.2\% of the time in S2,
followed by influencers at 23.7\%, and adopters contributing for 39.4\% 
to the total retweets. Opinion leaders, in this case as well,
emerge as the group with the least impact on the spread of information.

\section{Discussion}
\label{discussion}
The rise of social media has fundamentally transformed how
information, ideas, and opinions diffuse in contemporary society.
Whether traditional models designed in the social sciences to
understand this phenomenon, developed in the context of legacy
media forms in the middle of the twentieth century, still apply,
or whether we need a completely different way of thinking about
the issue remains a live question. 

We reconstructed the multi-step flow of information on a major social
media platform (Twitter) at the height of its influence. Our analysis
shows that the current structure of information diffusion in the social
media era displays characteristics of multiple combined processes. 
These include both unmediated one-step flows where users connect
directly to authoritative information sources, accounting for a small
percentage of the total activity. Mediated flows are far from
insignificant, particularly those that go via influencers, whether
of the traditional opinion leader or the more novel social media influencer kind.
This means that traditional two-step models still help make sense of this dynamic,
indicating that a lot of information flow in social media platforms is curated
and mediated by key actors, intervening between sources and potential adopters.
Longer multi-step flows can also be observed. Finally, horizontal
information flows among adopters, partially independent from
accredited or social-media-based mediators,
also account for a significant portion of the information flow. This dynamic
may be unique to the flow of information in digital platforms. 
The structure of information flow in the current system is thus closer
to a mixed regime, displaying dynamics differing in length, forms
of intermediation, and the type of actors involved. We also observe
intriguing differences in the prevalence of different elements of this
hybrid regime across the right/left political divide. In particular,
we find that social-media-based influencers, instead of opinion leaders,
leave an increased footprint in shaping public opinion
regarding right-related news than the left
case at earlier steps (see Fig. \ref{fig:left_vs_right}). 

Our work settles the question regarding the death of
intermediation and the decline of mediation and opinion
leadership in the social media era. While novel ways of accessing
information and distinct pathways of mediated information diffusion
have indeed opened up, opinion leadership is far from irrelevant.
Nevertheless, traditional opinion leaders face significant competition
from actors whose source of influence is social media reach. 
Only opinion leaders who themselves adopt the influencer strategy
may be able to become significant intermediators in the social media-driven ecology.

It is worth noting that, by design, the backbone
resulting from the BFS algorithm is only an approximation of the
information diffusion network, intended to propose a structure 
through which information is most likely to diffuse. The steps
involved in constructing and analyzing this backbone rely on
aggregations and averages, aiming to reconstruct a plausible
proxy for the diffusion of political-related news.
However, this approach loses the temporal dimension,
which is a critical aspect that could reveal
interesting differences in diffusion times as
information passes through different types of mediators.
Moreover, it's important to consider that our analysis is
performed exclusively on Twitter data. As a result, the
findings may not necessarily be applicable to other social
platforms, which could have different network structures,
user behaviors, or content-sharing dynamics. Given the
increasing heterogeneity of social media systems
in the post-Twitter era, future work should adopt a
comparative strategy across the plethora of emerging
platforms to investigate whether intermediation
dynamics of information flow differ systematically,
both cross-platform and across language, cultures,
topics, and even political divisions, as we observe
in this study. The framework we develop in this paper
can be readily adapted and scaled for such comparative studies. 
It is possible that different platforms may encourage
their own unique signature combination of one-step, two-step,
and multistep information flows, including more horizontal information
flows in platforms less dominated by influencer and opinion-leadership dynamics.
A deeper understanding of how particular policies and design decisions of
different platforms shape the particular structure of
information flow within them can provide insights for developing strategies
for effective communication and information campaigns at scale.

In the middle of the twentieth century, scholars in sociology and mass communication
studies could imagine relatively simple two-step dynamics in which
opinion leaders controlled the flow of information to the general public.
We are unlikely to ever go back to that world. Nevertheless,
the core insight that information mediation is an important
phenomenon, one that can co-exist with other ways of both
accessing and learning about news ideas and opinions, is
one that will continue to be central in the era of social media.


\section{Materials and Methods}\label{MaterialandMethods}
\subsection{Data} We track the spread of political
news on Twitter in 2020 by analyzing the dataset containing tweets posted
between 1 June and election day (2 November 2020). The data were collected
continuously using the Twitter search API with the names of the two presidential
candidates as keywords. The 2020 dataset contains 702 million tweets sent by 20
million users \cite{bovet2018validation,bovet2019influence,zhou2021polls,
flamino2021shifting}. 

To control for information polarization  \cite{flamino2021shifting}, we consider
tweets containing at least one URL link directing to a news media outlet
in a curated list of media outlets. The news outlet classification relies
on the website \href{www.allsides.com}{all-sides.com} (AS, accessed on 7 January 2021). 
We classified URL links for outlets that mostly conform to professional
standards of fact-based journalism in five news media categories: right,
right-leaning, left-leaning, and left. The classifications ('left' and 'right')
of media outlets used are subjective and sourced from publicly available datasets
by fact-checking organizations. A detailed explanation of the methodologies
used by AS for rating news outlets is given in \cite{bovet2018validation,
bovet2019influence,zhou2021polls,flamino2021shifting}. The full lists of
outlets in each category can be found in Appendix, Table A1. These
news outlets represent the sources of information of the information
diffusion model. The dataset under study contains 72.7 million tweets
with news links from one of these news outlets sent
by 3.7 million users. 

\subsection{Retweet network}
\label{Retweet Network}

The initial phase of investigating the real-world system involves
defining the retweet network, serving as a schematic representation
of the diffusion of political opinions (Appendix, Fig. A1a). This network is constructed 
by considering retweets containing a URL leading to one of the news outlets
introduced above. Two users ($i$ and $j$) are connected if one
has retweeted the other at least once \cite{pei2014searching}. Link directions
follow the flow of information, with the link going from $i$ to $j$ if $j$ has
retweeted a tweet from $i$. The resulting network is both directed and weighted,
with weights denoted by the variable \textit{w}, representing the number of times
user $j$ has retweeted user $i$.
Furthermore, since each tweet is timestamped, we calculate
the average retweet time between two nodes.
Before proceeding, it is necessary to address critical considerations to set the stage for subsequent steps that we must take to operate on the original retweet network. The Twitter data do not allow us to directly construct the diffusion cascade. Consider this scenario: User 0 posts news, marked by a specific URL. User 1 retweets this post directly from User 0's tweet. Subsequently, User 2 retweets from User 1's post. Ideally, the data for User 2’s tweet should cite User 1 as the source. However, the system identifies User 0's original tweet as the source instead. This pattern repeats with subsequent users, resulting in each tweet pointing back to User 0’s original post, thereby creating a star graph. Consequently, the final retweet network, composed of multiple star graphs, fails to accurately represent the actual diffusion pathways. To address this, we have developed a strategy that involves a validation and mapping process designed to reconstruct an \enquote{average} cascade structure. This approach allows us to identify the most relevant configuration of the diffusion process.

\subsection{Links Validation}
\label{Validating the Connections}

This step aims to preserve only statistically significant
connections \cite{Serrano2009Backbone}. The presence of numerous
non-statistically significant links could compromise our results
when determining the optimal model for information diffusion. 
For instance, if many adopters have connections
with a source with a weight of one, while only a few connections exist
with opinion leaders with weights greater than one, considering all
connections might incorrectly suggest the one step model as the best
description of information flow. However, upon statistical validation,
the weight-one links would be eliminated, emphasizing the connection
to opinion leaders as the most significant one, favoring the two step model.
Validation ensures that the observed connections are not random but
are influenced by the shift in the communication paradigm defined by 
the multi step model. To validate the structure and assign statistical
significance to the observed multi-step model structure, we employ null
models. Using null models helps us determine whether a connection between
two nodes is unexpected, potentially introducing misleading information,
or whether it is expected, indicating a meaningful flow of information 
between the two nodes. Therefore, selecting the appropriate null model
is essential to test the properties considered relevant and to adequately
address the research question \cite{Gotelli2000}. Hence, the pertinent question
becomes: Given the network we are examining, is it typical for a node i with
an out-strength of \(s_{out}\) and a node j with an in-strength of \(s_{in}\)
to be connected? Here, \(s_{in}(i) = \sum_j w_{ij}\) and \(s_{out}(i) = \sum_{j} w_{ji}\).

To accomplish this task, we employ maximum entropy models, a versatile class
of models that can incorporate fluctuations in measurements \cite{Fluctuating_Networks_2022},
thereby enhancing pattern detection quality \cite{Bruno2020}. 
These models assume different expressions based on the specific constraints
to be reproduced. Although analytical solutions for these models are 
rarely available, significant progress has been made in addressing
this challenge. Various models have been developed, ranging from
those suited for bipartite networks \cite{saracco2015randomizing}
to time-varying graphs \cite{clemente2023temporal}. Specifically,
we leverage the Conditional Reconstruction Method, a maximum
entropy ensemble model \cite{parisi2020faster}, for its proficiency
in accurately replicating observed system topologies while permitting weight
randomization. By doing so, it evenly distributes weights
across all available links. Our objective is to determine whether the
observed weight of a connection significantly deviates from the average
predicted by the ensemble. If the observed weight is markedly lower, we 
may consider severing that link. We retain the in- and out-strengths during
randomization because these metrics could reflect the characteristics of the
nodes themselves rather than being inherently linked to the connection.
For instance, while some individuals might be more inclined to retweet or
be retweeted, we aim to preserve this information. However, we simultaneously
control whether the existence of a retweet to or from a specific individual
can be justified.

Moreover, beyond validating based on the weight of connections, we also investigate whether the final inferred shape of the network, obtained using the next step of our strategy, changes when we apply additional filtering based on the rapidity of retweets. The rationale is to determine if removing significantly slow links, presumed to result from retweets that occurred with considerable delays and, therefore, likely reached new users through a chain of intermediaries, might distort the link back to the original source of information. To this end, we apply two different filters to the validated networks: the first retains only links that occur within 75
In both scenarios tested, we observed no significant deviations between the structures with and without temporal filtering. This finding suggests that, for the primary purpose of mapping information diffusion by focusing on groups of nodes rather than individual elements, the Breadth-First Search (BFS) algorithm applied to the original validated networks effectively captures the top fastest connections. Based on these observations, our results represent the general case without the need for temporal filtering.

To summarize, the validation process, by mainly exploiting the weight of the connections (Appendix, Section A1), enables us to concentrate on edges that hold more information or significance within the retweet network, offering a more precise representation of the underlying structure through which information propagates.

\subsection{Influencers and Opinion Leaders Identification}
\label{Influencers and Opinion Leaders Identification}

To determine the most suitable information propagation
model for Twitter, we need to identify the actors
of the model (Appendix, Fig. A1c): sources, opinion leaders 
(sometimes referred to as traditional opinion leaders), 
influencers, and the overlap between opinion leaders
and influencers, termed opinion-leading influencers.

In order to identify opinion leaders, we examine the URL
field in the user's Twitter object. Journalists/reporters 
associated with a news outlet often include a link redirecting
to their outlet in their bio. Therefore, we consider users 
linking to one of the news outlets classified as sources as
opinion leaders. We also consider politicians' profile
as opinion leaders. This set of users undergoes manual verification.
Please refer to the section on Data, Materials, and Software
Availability for a comprehensive list of classified sources
and opinion leaders.

To identify influencers in the validated retweet network,
we use the CI algorithm \cite{morone2016,
flamino2021shifting,serafino2022digital} (See Appendix, Section B).
This widely recognized metric identifies nodes whose removal 
could disrupt the giant connected component, influencing information 
diffusion. We select the top 1000 individuals with the highest CI 
scores among users with non-zero CI values. The top 1000 influencers
alone account for more than 85\% of the interactions in the network.
To check for users indirectly associated with news outlets,
a secondary check is performed on the top 1000 influencers 
identified by CI. Each influencer is classified
as an opinion-leading influencer if is an opinion leader. 
Otherwise, the user is labeled as an influencer. Users 
not belonging to sources, opinion leaders, influencers,
or opinion-leading influencers are labeled as adopters.
These groups of users have an empty intersection by definition. 

As an example when examining the retweet network derived
from utilizing all sources independently, we observe that Donald J.
Trump (former US president), Joe Biden (current US president),
and Natasha Bertrand (CNN reporter) are identified as opinion-leading influencers. Similarly, Jonathan Landay (Reuters reporter)
and Rick Tyler (political analyst at MSNBC) are recognized as opinion
leaders, while Donald Trump Jr. and Eric Trump are categorized as influencers.
Refer to Appendix, Table A2 for details. See Table \ref{tab:description_actors} for a synthetic description of the user categories and how they are identified.
It is worth noting that an alternative approach for identifying
opinion leaders could have been to use the verification badge provided in
Twitter's bio information. However, we found that many journalists do not
have this verification badge, which would result in the exclusion of many
traditional opinion leaders. Additionally, recent changes to
Twitter's policies allow users to purchase verification badges,
making this distinction less reliable. Despite this, as mentioned earlier,
the top 1000 influencers (0.1\% of the total users) identified by CI
account for more than 65\% of the connections in the network. Since these users
were manually checked and labeled, we are confident that the most important actors,
in terms of network structure, are included in this list. In other words,
while we may miss some opinion leaders or influencers,
those omitted have minimal impact on the network structure and,
therefore, would not significantly affect our results.

\begin{table}
\centering
\begin{tabular}{ p{4cm} p{10cm} }
    \textbf{Category} & \textbf{Description} \\
    \hline
    Sources & Accounts directly linked to a curated list of news outlets (refer to Appendix Table A1). \\
    \hline
    Influencers & Users who are among the top 1000 most influential nodes in retweet networks, as determined by Collective Influence, but are not considered Opinion Leaders. \\
    \hline
    Opinion Leaders & Users recognized as experts or respected public figures with acknowledged credibility in specific fields, identified here as journalists or other political figures who directly link to one of the considered news outlets in their descriptions. \\
    \hline
    opinion-leading influencers & Opinion Leaders, as defined above, who are among the top 1000 most influential users according to the CI. Among them, there are also well-recognized politicians or figures that can be directly associated with a political orientation.  \\
    \hline
    Adopters & Users not included in any of the aforementioned categories. \\    
\end{tabular}
\caption{A brief description of each category considered in this study is provided. Accounts described as Opinion Leaders, Influencers, and opinion-leading influencers have been manually reviewed.}
\label{tab:description_actors}
\end{table}

\subsection{Mapping the Information Flow: Breadth First Search}
\label{BFS}

To identify the information
diffusion model that best characterizes the Twitter information 
diffusion network, we employ a Breadth-First Search (BFS) algorithm.
The exploration begins with users classified as sources, serving as the
root nodes in the Breadth-First Search (BFS) algorithm. 
We examine all the first neighbors of these sources, distinguishing
this set of users into influencers, opinion-leading influencers,
opinion leaders, and adopters. This initial step identifies
the first step (S1) of the information diffusion model (Appendix, Fig. A1d).
At this stage, we consider only connections among users with a weight above one.

Next, we consider all the first neighbors of the newly identified
nodes (the first neighbors of the first neighbors), making sure not
to select users already chosen in the previous step. This step
identifies the second step (S2) in the information diffusion process.
Each iteration of the above procedure adds a new step in the information 
diffusion process and our algorithm halts when no more neighbors for
the nodes defined in the earlier steps are available.

\subsection{Data, Materials, and Software Availability}
The Twitter data and codes can be accessed on the following link: \\ \href{https://osf.io/u9svz/}{https://osf.io/u9svz/}.

\section{Acknowledgment}

HAM was supported by National Science Foundation Grant NSF-SBE award 2214217.
BKS was partially supported by DARPA under contract HR001121C0165, and the National Science Foundation Grant NSF-SBE 2214216. OL was partially supported by National Science Foundation Grant NSF-SBE 2214216.

\clearpage
\newpage

\appendix
\section*{Appendix}

\section{Validation}

We adopt a reference model constructed specifically to validate the connections between nodes in our network. This model is designed to preserve the topology of the retweet network, as well as the expected value of the total number of retweets made and received by each node. The resolution to an analogous challenge is detailed in a study where the authors introduce the CReMa (Conditional Reconstruction Method Model A) \cite{parisi2020faster}. The CReMa model allows to define a probability distribution over a set of graphs, that effectively replicates both the topology and the expected values of the network's in and out strength sequences. Additionally, it ensures that all other network observables are maximally random, and can be either analytically or numerically calculated \cite{squartini2017maximum}. The main tool employed in defining the model relies on the fundamental principle of Entropy Maximization \cite{Jaynes_1957}.

In our case, we estimate node-specific parameters $\vec{\beta}_{in}$ and $\vec{\beta}_{out}$ that are intrinsic to the model, correlating directly with the count of incoming and outgoing retweets for each node. This process allows us to calculate the expected weight ($w_{ij}$) of each link (how many times node i has retweeted node j) as well as its standard deviation. This enables the definition of the probability of observing the actual weight, considering only node-specific characteristics and not those of the individual link.

Specifically, for the employed model, the expected value and the standard deviation are the same and have the following expression:

\begin{equation}
    \sigma(w_{ij}) = \langle w_{ij} \rangle = \frac{1}{(\beta_{in} (i)+\beta_{out}(j))},
\end{equation}
from here we can attach a z-score to each link \cite{Gotelli2000} that is:

\begin{equation}
z(w_{ij}) = \frac{w_{ij}^* - \langle w_{ij} \rangle }{\sigma(w_{ij})},
\end{equation}
The parameters used to define the probability of connection in the referenced null model are obtained by solving the following 2N set of coupled equations (where N is the number of nodes):

\begin{equation}
\left\{
\begin{aligned}
\sum_{j(\neq i)} \frac{a_{ij}^*}{\beta_i^{\text{out}} + \beta_j^{\text{in}}} = s_i^{\text{out}*}, \quad \forall \, i \in N \\
\sum_{j(\neq i)} \frac{a_{ji}^*}{\beta_j^{\text{out}} + \beta_i^{\text{in}}} = s_i^{\text{in}*}, \quad \forall \, i \in N
\end{aligned}
\right.
\end{equation}
where \( s_i^{\text{out}*} \) and \( s_i^{\text{in}*} \) represent, respectively, the total number of times node \( i \) has been retweeted and the number of times node \( i \) has retweeted. $a_{ij}^*$ represents the element the topology of the retweet networks, that in our case is kept fixed. We chose to randomize while keeping the network's topology fixed because, in this case, the topology of the retweet network crucially represents the propagation of information among users. Randomizing by introducing connections between users who have never retweeted each other would distort this structure. Therefore, we opted to preserve the integrity of the original connections.
For further details on the model and calculations details, see \cite{parisi2020faster}.

After calculating the z-score, it's possible to establish a threshold. With this threshold, we can assess whether the observed actual value can be accepted or rejected based on the subjective evaluations. In our analysis, we have chosen a threshold of -1. Therefore, we accept all links that have a z-score within the range [-1, +$\infty$].

\subsection{Temporal Filtering}

We introduce an additional step to incorporate the temporal dimension into our analysis. After validating the connections by assessing the frequency of retweets among users, this final step involves testing and applying a final filter based on retweet timing before executing the BFS algorithm. For each connected pair of nodes in the retweet network, we calculate an average retweet time from the time differences between the original tweet and its retweets, each timestamped. We suppose that if the average retweet time between two users, $i$ and $j$, is longer than that involving another user, $k$, it indicates that user $i$ typically accesses information from user $j$ before user $k$.
Building on this assumption, and given that the BFS algorithm in the subsequent step will consider only one connection per pair of users to deduce the most relevant diffusion pattern, we analyze the distribution of average retweet times. We then apply and evaluate two types of filters: one retaining links within the top 20\% of this distribution and another within the top 75\%. Links outside these thresholds are removed. The BFS algorithm is then applied to this refined network structure.

\begin{figure}[h!]
    \centering  \includegraphics[width=0.9\textwidth]{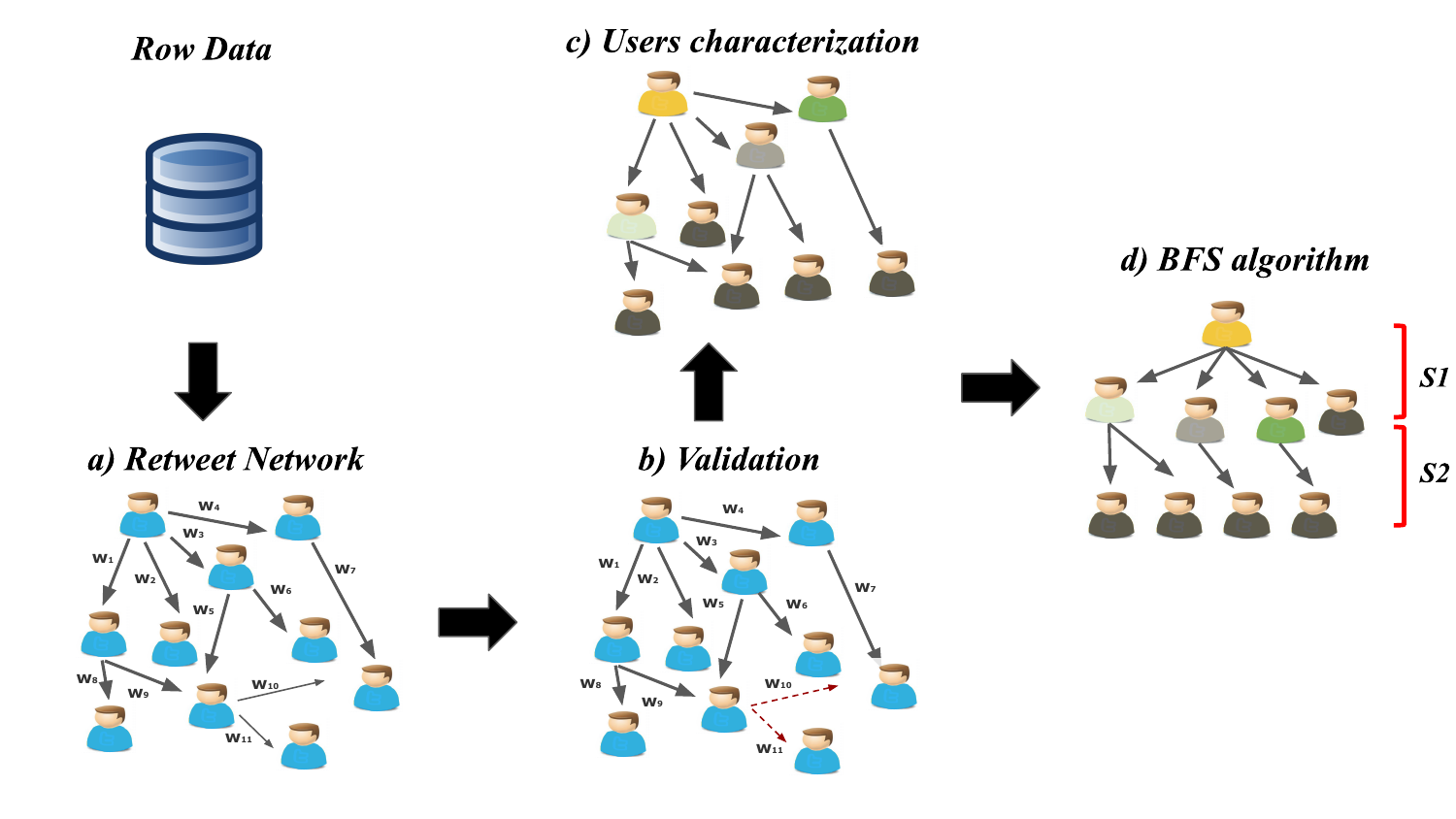}
    \caption{{\bf Pipeline}. Illustrative representation of the methodology 
    we follow in this work. (a) Starting from the raw data, we build a retweet network
    by considering all the retweets containing a URL redirecting to one of the news outlets
    in Table \ref{table:newsoutlets}. To each connection, we associate weights (w)
    and an average $\Delta t$ as explained in section \ref{MaterialandMethods}.
    (b) The retweet network undergoes link validation, and (c) after this step,
    users are classified as sources (sandy yellow), opinion leaders (green),
    opinion-leading influencers (light green), influencers (light gray),
    or adopters (dark gray). (d) On this network, we perform a BFS algorithm
    to identify the backbone (or skeleton) of the information diffusion.}
    \label{fig:pipeline}
\end{figure}

\begin{table}[ht!]
\center
\resizebox{.75\textwidth}{1.5in}{
\begin{tabular}{|l|r|r|r|l|}
\hline
& \multicolumn{2}{c|}{Left Leaning News} & \multicolumn{2}{c|}{Right Leaning News} \\
& hostnames & username & hostnames & username \\
\hline
1  & nytimes.com            & nytimes         & nypost.com             & nypost            \\
2  & washingtonpost.com     & washingtonpost  & wsj.com                & WSJ               \\
3  & cnn.com                & CNN             & forbes.com             & Forbes            \\
4  & politico.com           & politico        & washingtontimes.com    & WashTimes         \\
5  & nbcnews.com            & NBCNews         & foxbusiness.com        & FoxBusiness       \\
6  & theguardian.com        & guardian        & thebulwark.com         & BulwarkOnline     \\
7  & theatlantic.com        & TheAtlantic     & marketwatch.com        & MarketWatch       \\
8  & abcnews.go.com         & ABC             & realclearpolitics.com  & RealClearNews     \\
9  & npr.org                & NPR             & detroitnews.com        & detroitnews       \\
10 & bloomberg.com          & business        & dallasnews.com         & dallasnews        \\
11 & cbsnews.com            & CBSNews         & rasmussenreports.com   & Rasmussen\_Poll   \\
12 & cnbc.com               & CNBC            & chicagotribune.com     & chicagotribune    \\
13 & axios.com              & axios           & jpost.com              & Jerusalem\_Post   \\
14 & msn.com                & MSN             &                     &                 \\
15 & news.yahoo.com         & YahooNews       &                     &                 \\
16 & independent.co.uk      & Independent     &                     &                 \\
17 & latimes.com            & latimes         &                     &                 \\
18 & citizensforethics.org  & CREWcrew        &                     &                 \\
19 & buzzfeednews.com       & BuzzFeed        &                     &                 \\
\hline
\end{tabular}}
\center
\resizebox{.75\textwidth}{1.5in}{
\begin{tabular}{|l|r|r|r|r|}
\hline
& \multicolumn{2}{c|}{Right News} & \multicolumn{2}{c|}{Left News} \\
& hostnames & username & hostnames & username \\
\hline
1  & foxnews.com             & FoxNews           & rawstory.com              & RawStory \\
2  & dailycaller.com         & DailyCaller       & msnbc.com                 & MSNBC \\
3  & washingtonexaminer.com  & dcexaminer        & thedailybeast.com         & thedailybeast \\
4  & justthenews.com         & jsolomonReports   & huffpost.com              & HuffPost \\
5  & thefederalist.com       & FDRLST            & politicususa.com          & politicususa \\
6  & dailywire.com           & realDailyWire     & palmerreport.com          & PalmerReport \\
7  & theepochtimes.com       & EpochTimes        & motherjones.com           & MotherJones \\
8  & nationalreview.com      & NRO               & vox.com                   & voxdotcom \\
9  & saraacarter.com         & SaraCarterDC      & vanityfair.com            & VanityFair \\
10 & townhall.com            & townhallcom       & nymag.com                 & NYMag \\
11 & theblaze.com            & theblaze          & newyorker.com             & NewYorker \\
12 & thepostmillennial.com   & TPostMillennial   & dailykos.com              & dailykos \\
13 & westernjournal.com      & WestJournalism    & slate.com                 & Slate \\
14 & redstate.com            & RedState          & salon.com                 & Salon \\
15 & thegreggjarrett.com     & GreggJarrett      & rollingstone.com          & RollingStone \\
16 & bizpacreview.com        & BIZPACReview      & thenation.com             & thenation \\
17 & twitchy.com             & TwitchyTeam       & alternet.org              & AlterNet \\
18 & trendingpolitics.com    & CKeirns           & theintercept.com          & rdevro \\
19 & lifenews.com            & LifeNewsHQ        &                        &  \\
\hline
\end{tabular}}
\caption{ Hostnames in each media category. The tables contain information about the pages related to the news outlets considered in this study.}
\label{table:newsoutlets}
\end{table}

\begin{table}[ht]
\centering
\begin{minipage}{.5\linewidth}
\centering
\resizebox{\textwidth}{!}{
\begin{tabular}{|l|r|l|l|}
\hline
& \multicolumn{3}{c|}{Influencers} \\
\# & n$^\circ$ followers & name & username \\
\hline
1  & 664345    & The Lincoln Project   & ProjectLincoln \\
2  & 879979    & Laurence Tribe        & tribelaw \\
3  & 2248593   & Mark R. Levin         & marklevinshow \\
4  & 2428507   & James Woods           & RealJamesWoods \\
5  & 607927    & Tea Pain          & TeaPainUSA \\
6  & 5164374   & Donald Trump Jr.      & DonaldJTrumpJr \\
7  & 3685092   & Eric Trump            & EricTrump \\
8  & 1074520   & 60 Minutes            & 60Minutes \\
9  & 31896     & Don Moynihan          & donmoyn \\
10 & 109779     & Ryan Goodman        & rgoodlaw \\
\hline
\end{tabular}}
\end{minipage}%
\begin{minipage}{.5\linewidth}
\centering
\resizebox{\textwidth}{!}{
\begin{tabular}{|l|r|l|l|}
\hline
& \multicolumn{3}{c|}{opinion-leading influencers} \\
\# & n$^\circ$ followers & name & username  \\
\hline
1  & 81994886  & Donald J. Trump   & realDonaldTrump \\
2  & 638314    & Natasha Bertrand  & NatashaBertrand \\
3  & 1357350   & Maggie Haberman   & maggieNYT \\
4  & 6142647   & Joe Biden         & JoeBiden \\
5  & 687572    & Bill Kristol      & BillKristol \\
6  & 2564714   & Jake Tapper       & jaketapper \\
7  & 272043    & Greg Sargent      & ThePlumLineGS \\
8  & 804111    & Daniel Dale       & ddale8 \\
9  & 233827    & Jeffrey Goldberg  & JeffreyGoldberg \\
10 & 432294    & Peter Baker       & peterbakernyt \\
\hline
\end{tabular}}
\end{minipage}
\begin{minipage}{\linewidth}
\centering
\resizebox{.5\textwidth}{!}{
\begin{tabular}{|l|r|l|l|}
\hline
& \multicolumn{3}{c|}{Opinion leaders} \\
\# & n$^\circ$ followers & name & username  \\
\hline
1  & 36665   & Rick Tyler-Still Right & rickwtyler \\
2  & 17428   & Jonathan Landay        & JonathanLanday \\
3  & 34233   & jimrutenberg           & jimrutenberg \\
4  & 19490   & Henry J. Gomez         & HenryJGomez \\
5  & 8339    & Christian Datoc        & TocRadio \\
6  & 9655   & Michael Schwirtz           & mschwirtz \\
7  & 67721   & Charlie Savage         & charlie\_savage \\
8  & 132581  & Senator Ron Johnson    & SenRonJohnson \\
9  & 446379  & Sherrod Brown          & SenSherrodBrown \\
10 & 135322  & Leana Wen, M.D.        & DrLeanaWen \\
\hline
\end{tabular}}
\end{minipage}
\caption{ Example of influencers, opinion leaders, and opinion-leading influencers from the retweet network obtained by considering all the sources.}
\label{tab:example_INF_OL_OLI_FULL}
\end{table}

\section{Identifying Influencers}

We identify influencers using the Collective Influence (CI) method \cite{morone2015influence}, which involves an algorithm designed to find a minimal set of nodes capable of triggering a global cascade within the network, following the Linear Threshold Model \cite{kempe2003maximizing}.
For each node $i$, the CI is defined as follows:

\begin{equation}
\mathrm{CI}_{\ell}(i) = (k_i - 1) \sum_{j \in \partial \mathrm{Ball}(i, \ell)} (k_j - 1),
\end{equation}
where $\mathrm{Ball}(i, \ell)$ is the set of nodes inside a ball of radius $\ell$ around node $i$, with the radius defined as the shortest path distance, and $\partial \mathrm{Ball}(i, \ell)$ is the frontier (surface) of the ball. Here, $k_i$ is the degree of node $i$.
The value obtained for each node effectively evaluates the node's influence, considering the connectivity of nodes in its neighborhood. For our case, we choose $\ell = 1$.
Moreover, since this task is nondeterministic polynomial-time (NP) complete, the algorithm is impractically slow. Therefore, we apply a computationally efficient CI heuristic that provides an approximate solution.

\begin{figure}[!h]
    \centering 
    \includegraphics[width=170mm, height=100mm]{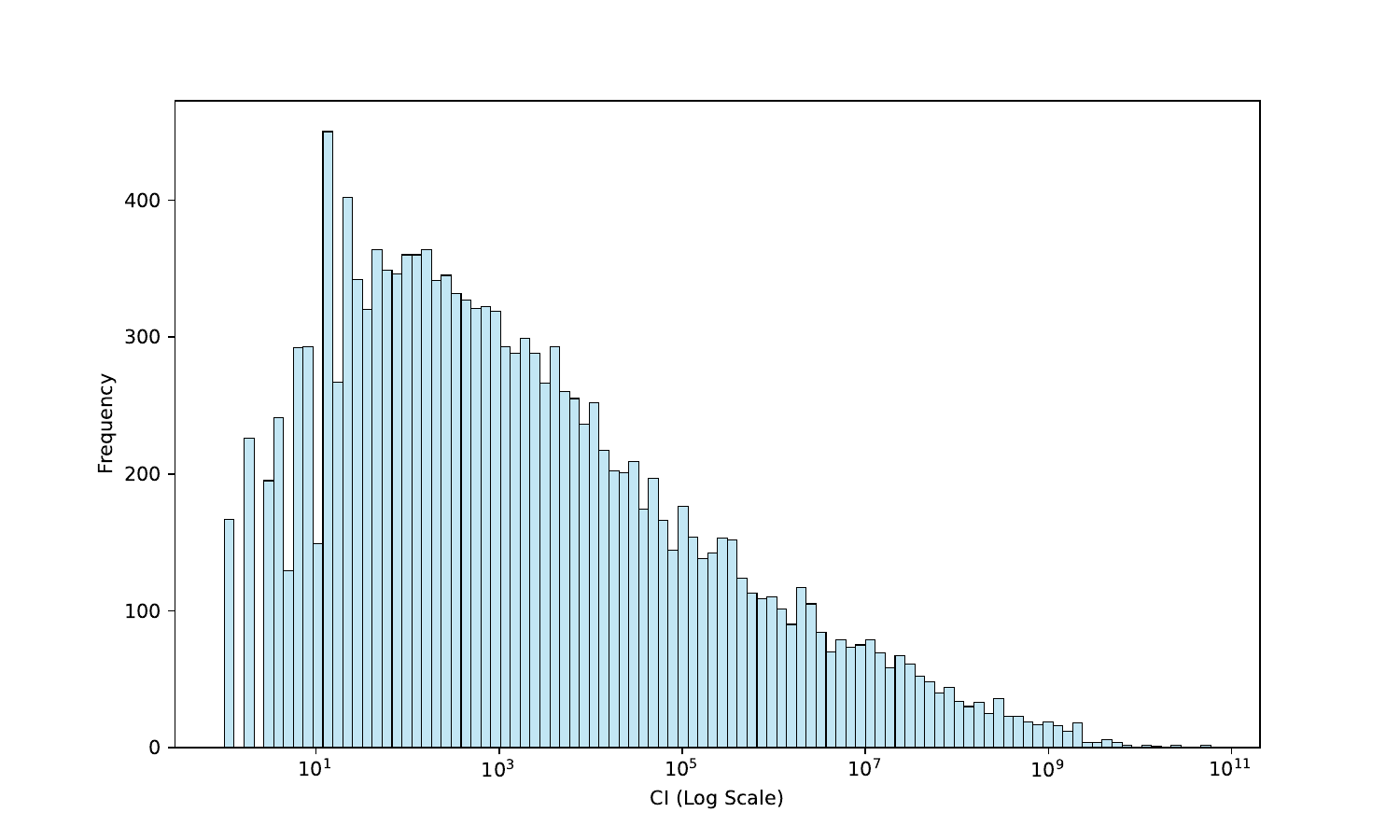}
    \caption{Distribution of Collective Influence in the Retweet Network, Analyzing News Related to Both Left- and Right-Leaning Sources.}
    \label{fig:distr_CI}
\end{figure}

After computing the CI for each node in the network, whose distribution for the general case network is represented in Fig. (\ref{fig:distr_CI}), we identify the most influential nodes. We select a number that captures, on average, the top $0.1\%$ of the influencers for the three categories studied: general case, Left, and Right; this resulted in an arbitrary threshold of 1000 elements as influencers.

\clearpage
\section{Further Analysis}
To support the results obtained using the BFS and to assess the potential loss of connections, and by extension, information, we measured the number of connections from Adopters to other types of actors within the validated networks. The idea is to assess how many connections go against he direction individuated by the BFS and, therefore observe the impact of or approximation.
The results are reported as follows:

\begin{table}[h!]
\centering
\begin{tabular}{|l|c|c|c|}
\hline
\textbf{Link Type} & \textbf{Left (\%)} & \textbf{Right (\%)} & \textbf{Full (\%)} \\ \hline
\textit{from} OL \textit{to} ADP & 0.0926 & 0.0555 & 0.0559 \\ \hline
\textit{from} I \textit{to} ADP & 0.1531 & 0.4571 & 0.0565 \\ \hline
\textit{from} OLI \textit{to} ADP & 0.0949 & 0.0310 & 0.0397 \\ \hline
\end{tabular}
\caption{Links pointing to ADPs from different types}
\label{tab:adp_links}
\end{table}
The low values observed in Table \ref{tab:adp_links} support our main results and corroborate the assumption that connections in other directions are less relevant to our primary hypothesis. 
These connections can be viewed in two ways: 
A part of those can be interpreted in relation to the results obtained in the BFS as the connections that goes from Adopters to the other categories in the Step 2  of the multi step model.
Another part of these connections can be viewed as noise on the main structure, which is obtained by averaging over many different events. The observation of such connections indicates that, although rare, there are instances where information flows in the opposite direction from that identified by our approach. However, these events are infrequent enough to be disregarded for the purposes of our analysis. 
Along with the robustness checks that incorporate temporal information, these results confirm the reliability of the inferred structure.

\clearpage

\bibliography{bibliography}

\end{document}